# Comment on: Testing the speed of 'spooky action at a distance'

Johannes Kofler, Rupert Ursin, Časlav Brukner, Anton Zeilinger

*Faculty of Physics, University of Vienna, Boltzmanngasse 5, A-1090 Vienna, Austria*
*Institute for Quantum Optics and Quantum Information (IQOQI), Austrian Academy of Sciences, Boltzmanngasse 3, A-1090 Vienna, Austria*

In a recent experiment, Salart *et al.*[1] addressed the important issues of the speed of hypothetical communication and of reference frames in Bell-type experiments. The authors report that they "performed a Bell experiment using entangled photons" and conclude from their experimental results that "to maintain an explanation based on spooky action at a distance we would have to assume that the spooky action propagates at speeds even greater than the bounds obtained in our experiment", exceeding the speed of light by orders of magnitude. Here we show that, analyzing the experimental procedure, explanations with subluminal or even no communication at all exist for the experiment.

In order to explain the violation of Bell inequalities within the view where, to use the author's wording, "correlated events have some common causes in their shared history", one needs to assume hypothetical communication between the observer stations. This communication must be faster than light if the outcome at one station is space-like separated from all relevant events at the other station.

In the experiment pairs of time-bin entangled photons were sent over 17.5 km optical fibers to two receiving stations, located in Jussy and Satigny, both equipped with a Franson-type interferometer and detectors. The *outcomes* were observed space-like separated from each other. The phase in the interferometer, i.e. the setting, in Jussy was continuously scanned, while the *setting* at Satigny was kept *stable*.

However, if the setting at one side remains unchanged, the results at both observer stations can be described by a "common-cause" without having to invoke any communication at all, let alone superluminal spooky action at a distance. This is signified, e.g., by the fact that no formulation of a bipartite Bell type inequality exists which does not use at least two settings at each side. Therefore, contrary to the claim in the paper, no Bell test was performed.

Furthermore, had the experiment been repeated with a second stable setting at Satigny, a "common-cause" explanation would still be possible. This is because in order to exclude subluminal communication, it is crucial that the *outcome event* on each side is space-like separated from the *setting choice* on the other side – which was not done in Ref. [1]. Thus, such experimental data – even if they were taken with two measurement settings at Satigny and even granting the fair-sampling assumption – could be explained by a "common-cause" model. In other words, the experiment tests the superluminal speed of hypothetical influences between outcome events under the assumption of no, not even subluminal, hypothetical influences between setting choices and outcome events.

We also remark that in a Franson-type experiment like the one reported in Ref. [1] the considered Clauser-Horne-Shimony-Holt Bell inequality is not applicable even with perfect detectors because of the inherent postselection.[2] One would (i) have to use a chained Bell inequality[2], (ii) achieve fast switching with a rate depending on the geometry of the interferometer, and (iii) reach a better visibility than the one reported in Ref. [1]. None of these three issues is covered by the experiment.

We would like to stress that this comment should not be seen as a defence of local realism. And neither do we demand that Ref. [1] must present a loophole-free Bell test. However, it is the purpose of our comment to point out "common-cause" explanations of an experiment which aims at putting "stringent experimental bounds on the speed of all such hypothetical influences".